\documentclass[num-refs,twocolumn]{wiley-article}

\usepackage{graphicx, amsmath,amssymb}

\usepackage{siunitx}
\usepackage{siunitx}

\papertype{Original Article}
\paperfield{Journal Section}

\title{Time-multiplexed Reservoir Computing with Quantum-Dot Lasers: Does more complexity lead to better performance?}

\author[1]{Huifang Dong}
\author[1]{Lina Jaurigue}
\author[1]{Kathy Lüdge*}

\affil[1]{Institut für Physik, Technische Universität Ilmenau, Weimarer Str. 25, 98693 Ilmenau, Germany}

\corraddress{Kathy Lüdge,Institut für Physik, Technische Universität Ilmenau, Weimarer Str. 25, 98693 Ilmenau, Germany}
\corremail{kathy.luedge@tu-ilmenau.de}

\runningauthor{Dong et al.}

\begin{document}

\begin{frontmatter}
\maketitle
\begin{abstract}
Reservoir computing with optical devices offers an energy-efficient approach for time-series forecasting. Quantum dot lasers with feedback are modelled in this paper to explore the extent to which increased complexity in the charge carrier dynamics within the nanostructured semiconductor can enhance the prediction performance.  By tuning the scattering interactions, the laser's dynamics and response time can be finely adjusted, allowing for a systematic investigation. It is found that both system response time and task requirements need to be considered to find optimal operation conditions. Further,   lasers with pronounced relaxation oscillations outperform those with strongly damped dynamics, even if the underlying charge carrier dynamics is more complex. This demonstrates that optimal reservoir computing performance relies not only on internal complexity but also on the effective utilization of these dynamics through the output sampling process.
\end{abstract}

\end{frontmatter}


\section{Introduction}\label{sec:level1}
The pursuit of efficient and adaptable computational frameworks has driven researchers to explore unconventional computing paradigms inspired by the brain \cite{ZOL24}. Among these is the reservoir computing scheme, which was originally introduced as liquid state machines \cite{MAA02} and echo state networks \cite{JAE01}, and was later unified under the name reservoir computing (RC) \cite{LUK09}. RC has garnered considerable attention due to its easily implementable architecture, where only the readout layer needs to be trained with a linear regression algorithm \cite{JAE01,MAA02}. This makes RC particularly attractive as it can greatly reduce computational costs and enhance hardware implementation \cite{SAN17a,TAN19c}. Recent reviews further highlight technological developments and future perspectives in this field \cite{ABD24,ZHA23d}.

The underlying recurrent network of a RC setup enables short-term memory, making the concept particularly promising for time-series prediction tasks \cite{WYF10, CAN18, MOO19, LYM19, GUO24}, as well as other applications that require temporal memory. These applications include action recognition \cite{PIC23, ANT19}, speech recognition \cite{PAQ12, LAR17}, signal classification \cite{ESC14}, and channel equalization \cite{BRU13a, ORT15, SKO22, ARG20, GUO24}. Although RC has also been utilized for pattern recognition \cite{JAL18, TON18}, in these cases, the temporal memory properties cannot be fully exploited.

Using photonic hardware for RC applications enables very fast and on-chip-compatible implementations \cite{BRU19}. This approach also allows for the exploitation of multiplexing techniques in both frequency \cite{DEL24, GUO24, FIS24} and space \cite{SKA22b, POR21, ANT19, PFL24}. For a comprehensive overview of these advancements, the reader is referred to recent reviews \cite{BAI23, ABR24}. In this paper, we focus on time-delayed RC, which was first introduced in 2011 \cite{APP11} and requires only one physical node. This approach is based on the concept of time-multiplexing and has proven to be a highly effective method for hardware implementations \cite{ABR24, BRU19}. When a semiconductor laser is utilized as the physical node, data can be easily injected into the reservoir through a pump current, while the readout process simply involves sampling the intensity over time at the laser's output facet. We will focus on quantum dot (QD) lasers, which have garnered significant interest due to their unique properties, such as precise control of the emission wavelength, lower threshold current, high-temperature stability, and compatibility with chip technology \cite{SHA21a}. Additionally, QD lasers introduce further degrees of freedom, thereby increasing the complexity of the system’s internal dynamics. Our aim is to clarify whether this heightened complexity can be leveraged to enhance RC performance. 

In the past, QD lasers have been explored as candidates for coherent optical chaotic secure communication systems \cite{ZHO24, ZHO23c}, RC systems \cite{HEU20, PFL24, TAN22, TAT24}, and frequency multiplexed computing applications \cite{SKO22}. Due to variations in material properties and size distribution, the modulation response of QD lasers can vary significantly based on the characteristic scattering timescales \cite{LIN12, LUE11a, LUE11, PAU12, LIN14}. By systematically tuning the effective scattering rates, we clarify how the dynamic properties can be exploited for RC. Surprisingly, we find that the higher dimensional dynamics of QD lasers do not contribute to better results. Instead, it is the more pronounced relaxation oscillations (ROs), akin to those observed in conventional quantum well (QW) lasers \cite{ERN10b, COL12a}, that enhance computing performance.

\begin{figure*}
\includegraphics[width=0.9\textwidth]{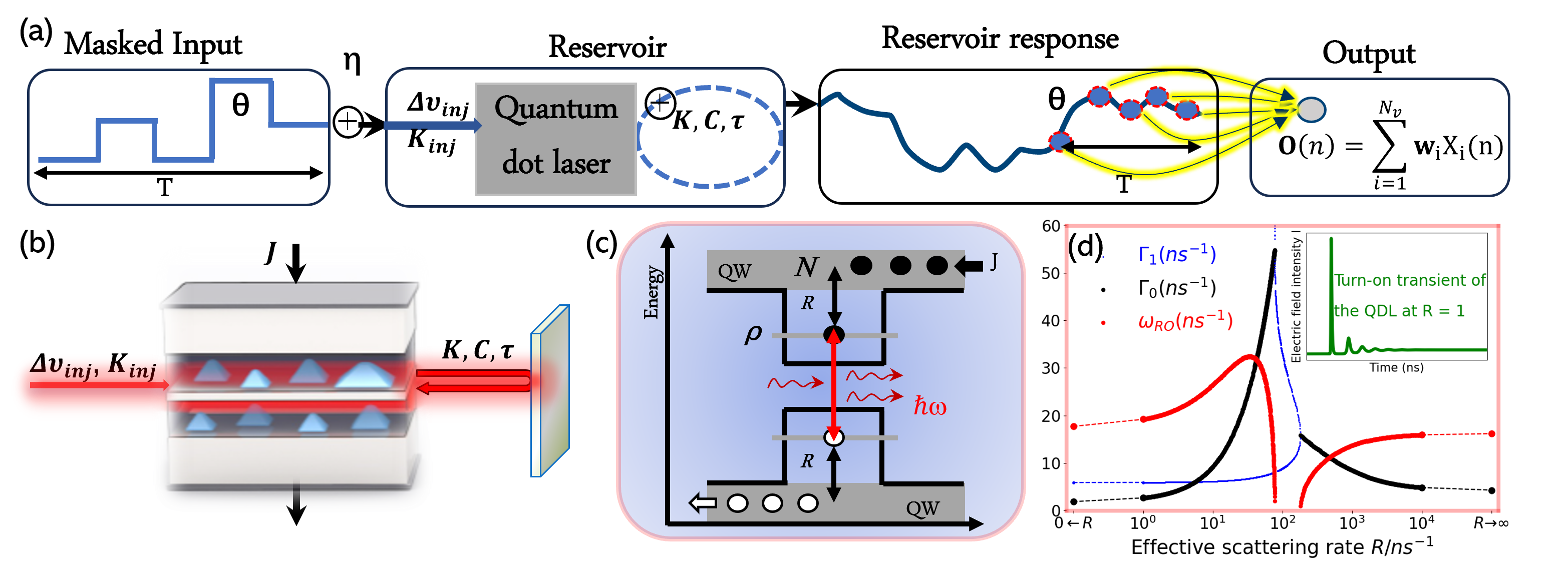} \caption{(a) Scheme of the QD laser RC with time$-$delayed feedback and external optical injection, masked inputs are fed into the reservoir, and all the responses are trained with linear weight $\textbf{w}$. (b) Sketch of the QD laser with time$-$delayed feedback and external optical injection reservoir. (c) Band structure diagram of the solitary QD laser. (d) RO angular frequency $\omega_{RO}$ (red), damping rate $\Gamma_{0}$ (black), and damping rate $\Gamma_{1}$ (blue), obtained from the solutions of Eq.(\ref{eq:character}),  as a function of the effective scattering rate $R$, and the insert shows the turn$-$on transient of the QD laser with $R=1$. Parameters are given in Tab.\ref{Tab:QDparameters}. \label{fig: figure1}}
\end{figure*}

We also discuss in detail the intricate relationship between input timescales and the system's response time, and explore the role of the ratio between the delay time and the input clock cycle.  While there is general agreement in the literature that it is beneficial to avoid integer multiples of delay time and clock cycle  \cite{JAC22, DEJ14, NAK16, ROE19, HUE22}, mainly due to their detrimental impact on the memory properties of RC systems \cite{KOE21, GOL20, KOE20a}, many researchers still utilize resonant setups \cite{SUG20, GUO20a, BAI23a}. Our analysis demonstrates that the optimal ratio between clock cycle and delay time is influenced by several factors, including the system's response time, the number of virtual nodes, and the memory requirements of the task at hand.


For our numerical simulations, we employ a simplified QD laser model introduced by Lingnau et al. \cite{lin15a} to explore how the computing performance varies as a function of the scattering lifetimes for two prediction tasks$:$ the Lorenz 63 and Mackey-Glass systems. Additionally, we investigate the effect of external optical injection on the system's performance. The article is structured as follows: The time-multiplexed RC setup, the QD laser model, the prediction tasks and error measures are introduced in Sec.\ref{sec:methods}. The results regarding the time series prediction performance of the different QD lasers for the two different tasks both with only self-feedback and with self-feedback and optical injection are presented in Sec.\ref{sec:results} before we conclude in Sec.\ref{sec:conclusion}.

\section{Methods and Model} \label{sec:methods}
\subsection{Time-multiplexing}\label{sec:level2}

The time-multiplexed RC setup is illustrated in Fig.\ref{fig: figure1}a. In this scheme, the physical node receives time-series data, which is multiplied by a mask to diversify the response over time. Subsequently, the masked input data are fed into the reservoir via modulation of the laser's pump current, where they are processed non-linearly by the internal dynamics of the QD laser. The reservoir is build by a QD laser that is subjected to both optical feedback and optical injection (Fig.\ref{fig: figure1}b). The laser response to one masked input step is measured by sampling the output intensity over time during one mask length!!!!. The computing output for one input is then obtained by multiplying the sampled output vector with the trained linear weight vector $\textbf{w}$. 

The masked input $I_{m}(i)$ is given by
\begin{equation}
I_{m}(i)=I(k)m(j)
\end{equation}
where $I(k)$ is the input signal with the length of $K_{tr}$, $m(j)$ is the mask with the length of $N_{v}$, which is the number of virtual nodes. The index of the masked input is thus $i=k N_{v}+j$, $k \in [0, K_{tr})$, $j \in [0, N_{v})$. By holding each masked input fixed for the virtual node interval $\theta$, the discrete input sequence transforms into a piecewise constant function:
\begin{equation}\label{eq:input}
J_{m}(t)=I(k)m(j) \ (k=[ t/(N_{v} \theta) ], j= [ t/\theta ] \textrm{mod} N_{v})
\end{equation}

In the training phase, $N_{v} \cdot K_{tr}$ masked inputs are sequentially fed into the reservoir, and the sequential responses of the RC are sampled every $\theta$ interval, corresponding to the piecewise constant input time series. All the responses are mapped into a state matrix $\textbf{S} \in \mathbb{R}^{K_{tr}} \times \mathbb{R}^{N_{v}+1}$ \cite{HUE22,JAU21a,ROE19}, the last column of $\textbf{S}$ is filled with a bias term of 1. The prediction output \textbf{o} of the RC is given by:
\begin{equation}
\textbf{o}=\textbf{S} \textbf{w}.
\end{equation}
Where the weight \textbf{w} is trained with the optimization of the problem:

\begin{equation}
\min \limits_{\textbf{w}} (||\textbf{S}\textbf{w}-\textbf{\^{o}}||_{2}^{2}+\lambda_{T}||\textbf{w}||_{2}^{2}).
\end{equation}
Here \textbf{\^{o}} is the target output, $\lambda_{T}$ is the Tikhonov regularisation parameter, and $|| \cdot ||_{2}$ is the Euclidean norm. 
The solution to this problem can be obtained with Moore-Penrose pseudoinverse \cite{BRU18a}:  
\begin{equation}
\textbf{w}=(\textbf{S}^{T} 
\textbf{S}+\lambda_{T}\cdot \textbf{I})
^{-1} \textbf{S}^{T} \textbf{\^{o}}
\end{equation}
where $\textbf{I} \in \mathbb{R}^{N_{v}+1} \times \mathbb{R}^{N_{v}+1}$ is the identity matrix.

\subsection{Solitary quantum dot laser}\label{sec:level3}

We model the dynamics of the QD laser using a minimal QD laser model \cite{lin15a}. This model is derived from microscopically-based rate equations designed specifically for the dot-in-well laser system, where all the Auger scattering processes between the electrons in the different bands are calculated ab-initio \cite{LUE09, LUE11a}. It is then simplified by assuming only one carrier type, linearization around the value at the laser threshold \cite{RED17}, and constant amplitude-phase coupling. Fig.\ref{fig: figure1}b illustrates the schematic of an $InAs/GaAs$ dot$-$in$-$well single mode QD laser device, comprising multiple GaAs QW layers, each layer randomly embedded with a density of $N^{QD}$ InAs QDs, and all these layers are encased by bulk semiconductor material. The energy band structure sketch of the QD laser is depicted in Fig.\ref{fig: figure1}c. 
The resulting rate equations are given by:
\begin{align}
 \begin{aligned}\label{eq:QDL_function}
\dot{N}&=J+J_{m}(t)\frac{\eta}{T_{1}}-\frac{N}{T_{1}}-R(\rho_{th}-\rho+d(N-N_{th})) \\ 
\dot{\rho}&=R(\rho_{th}-\rho+d(N-N_{th}))\\
&~~~~~~~~~~~-\frac{\rho}{T_{sp}}-[2 g(\rho-\rho_{th})+\kappa] |E|^{2}  \\
\dot{E}&=2 g(1-i \alpha)(\rho-\rho_{th})E  
\end{aligned}
\end{align}
where $E$ is the complex slowly varying amplitude of the electric field in the cavity, $N$ represents the total charge carrier (electrons and holes) density in units of $2N^{QD}$ in each QW layer,  and $\rho$ describes the occupation probability of the charge carriers in the QD. In addition, the $J_{m}(t)$ term is used for encoding information, while $\eta$ is the scale parameter of the input ($\eta=0$ describes the solitary QD laser).

Carriers are pumped into the QW by a pump current $J$. Stimulated emission is determined by both the gain coefficient $g$ and the population inversion $2(\rho-\rho_{th})$, where $\rho_{th}=(\kappa+g)/(2g)$ is the occupation probability of the QD levels at the laser threshold. $\alpha$ denotes the linewidth enhancement factor. This value is usually not a constant for QD lasers but dynamically depends on the carrier occupation in the non-resonant energy bands \cite{LIN12b}. This becomes important for large changes in the operation conditions \cite{MEI17, LIN17a}, which is not the case here. The coupling between $\rho$ and $N$ is expressed via the effective scattering term $R(\rho_{th}-\rho+d(N-N_{th}))$, where $R$ is the effective scattering rate, $N_{th}$ is the charge carrier density in the QW at the laser threshold and $d$ signifies the detailed balance coefficient \cite{lin15a}. Additionally, $T_{1}$, $T_{sp}$, and $\kappa$ denote the charge carrier lifetime in the QW, the spontaneous emission loss time in the QD, and the optical loss rate inside the cavity, respectively. The specific lasing wavelength is controlled by the size, distribution, and Indium concentration of the QDs. While the parameters chosen in this paper are specifically tailored to model $As$-based QD lasers, they can also be adjusted to describe QD lasers made from other material systems, e.g. InAs/InP-based QD lasers emitting at the telecommunication wavelength of $1550 \ nm$.

The turn-on properties and modulation response characteristics of QD lasers have been extensively studied in the past \cite{ERN07a, LUE09, LUE13, LIN10, LIN17a}. A unique feature of these lasers is that their turn-on transients change drastically with the charge carrier lifetimes, which are represented by the effective scattering rate 
$R$ in our minimal model. Tuning $R$ leads to QD lasers with strongly damped ROs, which resemble the characteristics of a Class-A laser \cite{ARE91}, as well as pronounced Class-B-like oscillations at very large scattering rates \cite{PAU12}. We analyze the RO damping rates and angular frequencies of the solitary QD laser via a linear stability analysis \cite{lin15a} of Eq.(\ref{eq:QDL_function}) around its fixed points. The eigenvalues $\lambda$ are given by the following characteristic equation$:$
\begin{align}
\begin{aligned}\label{eq:character}
0&=-\lambda(\frac{1}{T_{1}}+dR+\lambda)(R+\frac{1}{T_{sp}}+2gI^{*}+\lambda)\\
&-4g \kappa I^{*}(\frac{1}{T_{1}}+dR+\lambda)+R^{2}d \lambda \\
\end{aligned}
\end{align}
where $I^{*}=\frac{Rd(J T_{1}-N_{th})}{\kappa(1+Rd T_{1})}-\frac{\kappa+g}{2 g \kappa T_{sp} }$ is the optical intensity at the fixed point. We obtain a complex conjugate pair of eigenvalues $\lambda_{0,1}=-\Gamma_{0}\pm i \omega_{RO}$ and a single real eigenvalue $\lambda_{2}=-\Gamma_{1}$, where $\omega_{RO}$ is the RO angular frequency. The RO damping rate $\Gamma_{RO}=f(\Gamma_{0},\Gamma_{1})$ is given by the two real values $\Gamma_{0}$ and $\Gamma_{1}$. Both of them, along with $\omega_{RO}$ are illustrated in Fig.\ref{fig: figure1}d as a function of the effective scattering rate $R$. The inset shows one example of a turn-on transient for small $R$. 

In the limit $R \rightarrow \infty$, where all electrons instantaneously enter the lasing level, the three rate equations in Eq.(\ref{eq:QDL_function}) can be reduced to two, which resemble the equations for a semiconductor laser with a single carrier type, e.g. a quantum well laser \cite{lin15a, OTT11}. The limit $R \rightarrow 0$ also allows for a reduction to two equations, which is however only a mathematical limit rather than a physically realizable scenario as in that case no electrons will enter the lasing level. The equations for these limiting cases are provided in the supplementary material, the corresponding eigenvalues are illustrated in Fig.\ref{fig: figure1}d.

\begin{table}[ht]
\caption{Parameters of solitary QD laser} 
\begin{threeparttable}
\begin{tabular}{c c c} 
\headrow
Symbol \hspace{-1mm} & Value & Meaning \hspace{-1mm}\\ 
$J/J_{th}$ & 2 & current in units of threshold  \\
$T_{1}$ & 0.17ns & charge carrier lifetime in QW    \\
$\rho_{th}$ & 0.609 & threshold QDoccupation prob.\\
$N_{th}$ & 2.3 &  threshold QW carrier density  \\
& & (in units of $2N^{QD}$)\\
$T_{sp}$ & 1.85\,ns & spontaneous emission lifetime \\
$\alpha$ & 1.42 &  linewidth enhancement factor \\
$\kappa$ & $50ns^{-1}$ &  optical losses rate in cavity \\
g & $230ns^{-1}$ & gain coefficient\\
d & 0.022 & detailed balance coefficient\\
$\tau$ & 1.69\,ns & feedback cavity round-trip time\\
\hline 
\end{tabular}
\end{threeparttable}
\label{Tab:QDparameters} 
\end{table}

\subsection{Reservoir}\label{sec:level4}


The reservoir used for the time-series prediction task is the QD laser described in Eq.\eqref{eq:QDL_function}, initially subjected to optical self-feedback, and later extended to incorporate optical feedback and injection. It is well known that time-delayed feedback and external optical injection can not only significantly enrich the underlying bifurcation structure of laser systems \cite{ERN00, ERN10b, WIE05}, but also improve the reservoir computing performance \cite{MUE24, EST23, EST20, YAN22, ZHA22a}. Time-delayed feedback reintroduces the electric field from the past into the current state, enriching the reservoir's memory, while optical injection stabilizes the laser and increases its optical bandwidth, enabling better handling of complex and high-frequency input signals \cite{EST23, EST20}. We model both by adding terms representing time-delayed feedback and external optical injection to the differential equation governing the dynamics of the electric field. The modified electric field equation is as follows \cite{KOE21b, MUE24}: 
\begin{equation}
\begin{aligned}\label{eq:QDL_injected_feedback}
\dot{E}&=2 g(1-i \alpha)(\rho-\rho_{th})E+K  \kappa  e^{-i C}E(t-\tau)\\
&+K_{inj}  \kappa  E_{0}e^{-i (\omega_{0}+2 \pi \Delta \nu_{inj} )t},\\
\end{aligned}
\end{equation}
where $K$ is the feedback strength, $C$ is the feedback phase, $\tau$ is the delay time,
$K_{inj}$ is the injection strength, and $\Delta \nu_{inj}$ is the frequency detuning between the external injected driving laser and the free-running solitary QD laser. $E_{0}$ and $\omega_{0}$ are the steady-state values of the free-running solitary laser describing the complex electric field amplitude and its angular frequency. 







\subsection{Lorenz and Mackey Glass Tasks}\label{Tasks}
The Lorenz system, known for its rich chaotic dynamics, is a well explored benchmark task for timeseries prediction \cite{JAU24,JAU21a,HUE22a}. The equations are given by \cite{LOR63}:
\begin{align} \label{eq:Lorenz}
 \begin{aligned}
 \dot{x}&=c_{1}y-c_{1}x,  \\
 \dot{y}&=x(c_{2}-z)-y, \\
 \dot{z}&=xy-c_{3}z,
\end{aligned}
\end{align}
and with $c_{1}=10$, $c_{2}=28$, and $c_{3}=8/3$ chaotic dynamics is observed. We generate the Lorenz time-series using Runge-Kutta 4th order numerical integration with a time step of $h_{0}=10^{-3}$. Our input signal $I(k)$ is obtained via sampling the $x$ variable with a step size of $dt = 0.1$. The target for the prediction task is $I(k+\Delta s)$ with $\Delta s=1$, e.g. a one step ahead Lorenz x-to-x prediction. All inputs and targets are rescaled to the range $[0,1]$.


Another common benchmark task is the prediction of the Mackey-Glass time-series   \cite{JAU24,JAU21a,HUE22a}, which is generated by the Mackey$-$Glass delay-differential equation \cite{MAC77}$:$
\begin{equation}
\frac{dx}{dt}=\frac{\beta x(t-\tau_{0})}{1+x(t-\tau_{0})^{n}}-\gamma x
\end{equation}
with $\tau_{0}=17$, $n=10$, $\beta=0.2$, and $\gamma=0.1$. The Mackey-Glass time-series is generated with a time step of $h_{0}=10^{-2}$ using a 4th order Runge-Kutta method with Hermitian interpolation. The input signal $I(k)$ is sampled with a step of $dt=1$, and the corresponding target sequence is defined as $I(k+\Delta s)$. It is mentioned that changing $\Delta s$ or the sampling step $dt$ changes the task and its memory requirements \cite{JAU24, TSU23},  which will change the expected performance.  In this paper, we choose $\Delta s=2$ for a two-step-ahead prediction of this system as our target sequence. All inputs and targets are rescaled to the range $[0,1]$.

\subsection{Error measure}\label{Sec:error}

The prediction performance of the RC for a specific task is quantitatively evaluated with the normalized root-mean-square error (NRMSE), denoted as $\delta$:
\begin{equation}\label{eq:NRMSE}
\delta=\sqrt{\frac{ \sum_{k=1}^{K_{tr}} (\textbf{\^{o}}_{k}-\textbf{o}_{k})^{2}}{K_{tr}\textbf{var}(\textbf{o})}}
\end{equation}
where $k \in [0, K_{tr})$ is the index of the input data, and $K_{tr}$ is the length of the input data sequence, $\textbf{o}_{k}$ and $\textbf{\^{o}}_{k}$ are the target output and prediction output, respectively, and $\textbf{var}$ represents the variance.

In a single trial, $\delta$ is calculated using an identically distributed random mask for the reservoir. To mitigate the impact of outliers resulting from poorly performing masks, the prediction process is repeated over 30 trials. The median from these trials is denoted as $\widetilde{\delta}=\textrm{median}(\delta[i])$, where  $i \in [0, 30)$ is the index of each trial. To quantify the error variability, the error bar for $\widetilde{\delta}$ is determined using the median absolute deviation (MAD), defined as: $MAD=\textrm{median}(|\delta[i]-\widetilde{\delta}|)$. 

In addition, to assess how strongly the computing error $\widetilde{\delta}$ depends on the hyper-parameters of the RC setup, we employ a measure that evaluates both the best value and the robustness of the performance over a 2-dimensional parameter plane, as described in \cite{JAU24}. For a given parameter plane spanned by parameters A and B, we determine the minimum error $\widetilde{\delta}_{min}^{AB}$, which represents the smallest $\widetilde{\delta}$ value achieved within the $(A, B)$-plane, and the sensitivity $P_{AB<threshold}$, which quantifies the percentage of the $(A, B)$-plane where the $\widetilde{\delta}$ value is below a chosen threshold:
\begin{equation}\label{PKC}
P_{AB<threshold}=\frac{S_{AB<threshold}}{S_{AB}}
\end{equation}
where $S_{AB<threshold}$ represents the area of the $(A, B)$-plane where $\widetilde{\delta}$ is below the threshold, and $S_{AB}$  represents the total area. This sensitivity measure helps us evaluate the robustness of the system, indicating how dependent the computing performance is on the choice of the hyper-parameters. A higher value of $P_{AB<threshold}$ suggests that good performance can be achieved over a broader range of parameter settings, implying greater stability and reliability of the reservoir setup, which is advantageous in practical applications. Depending on the tasks we use different threshold values. For the Lorenz task with $\widetilde{\delta}_{min}^{AB}\approx 0.01$ we use $\widetilde{\delta}_{threshold}=0.032$, and for the Mackey-Glass task, with $\widetilde{\delta}_{min}^{AB}\approx0.001$, we use $\widetilde{\delta}_{threshold}=0.01$ if not noted other\-wise.  

\subsection{Simulation Conditions}
All simulations are performed in C++ and the Library 'Armadillo' is used for the linear algebra calculations. The delay-differential equations describing the reservoir are numerically integrated using a Runge$-$Kutta 4th order method with Hermitian interpolation and an integration step of $h = 0.001$. The reservoir is initialized with an input sequence of length 1000. Subsequently, the system is trained on $K_{tr} = 5500$ inputs. Following the training phase, another buffer of 1000 inputs is introduced before the testing phase performance is evaluated on a sequence of $K_{test}=5500$ inputs. The prediction performance of the RC is evaluated with the median error $\widetilde{\delta}$
over 30 trials with randomly selected input masks from a uniform distribution in the range $[-1,1]$. A Tikhonov regularisation parameter of $\lambda_{T}=10^{-9}$ is used in this paper. We do not use additional noise in the QD laser simulations.

\section{Results and discussion}\label{sec:results}


\subsection{QD laser dynamics with optical feedback}\label{sec:level5_1}

We first provide insights into the emission dynamics of the QD laser subjected to optical feedback, focusing particularly on the bifurcations in the system's dynamics. These bifurcations play a crucial role in influencing the computing performance of the QD laser reservoir. 

\begin{figure}[t]
\includegraphics[width=0.49\textwidth]{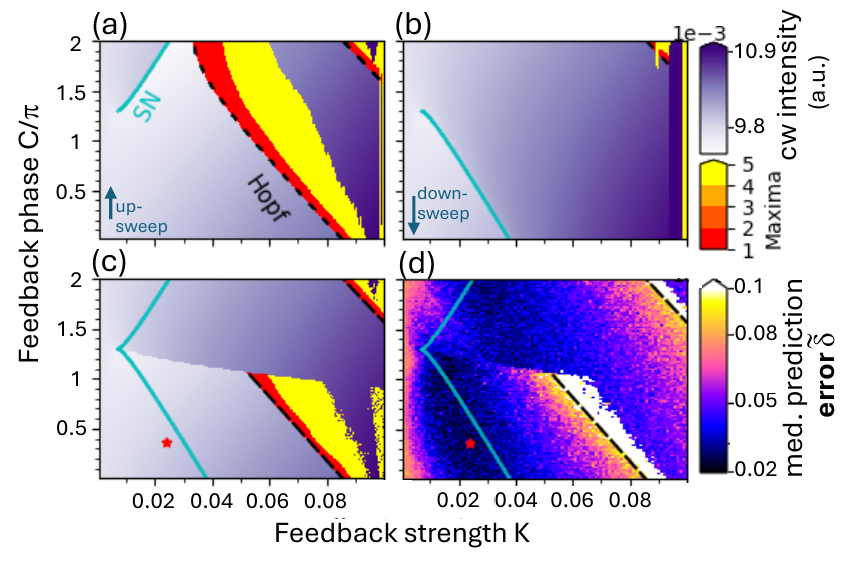}     \caption{2D emission dynamics in the $(K, C)$- plane for a QD laser with optical feedback ($R$=1, $\tau=1.692$\,ns): (a) initial conditions chosen via \textit{up-sweep} in $C$, (b) \textit{down-sweep}  in $C$, and (c) without sweep (constant initial cond.), arrows indicate sweep direction. Constant wave (cw) emission is encoded via blue shading while the number of unique maxima is encoded with red/yellow color. (d) Prediction performance $\tilde{\delta}$ (as defined in Eq.\eqref{eq:NRMSE}) for the Lorenz task (without sweep). Cyan lines represent saddle-node (SN) bifurcations (determined analytically from Eq.\eqref{eq:QDL_function}), black dashed lines indicate Hopf-bifurcations observed numerically. Red star marks best performance $\tilde{\delta}_{min}^{KC}$. Parameters: $N_V=30$ and as listed in Tab.\ref{Tab: CW-NRMSE} and \ref{Tab:QDparameters}.\label{fig: figurenew}}
\end{figure}

\begin{figure*}
\includegraphics[width=0.95\textwidth]{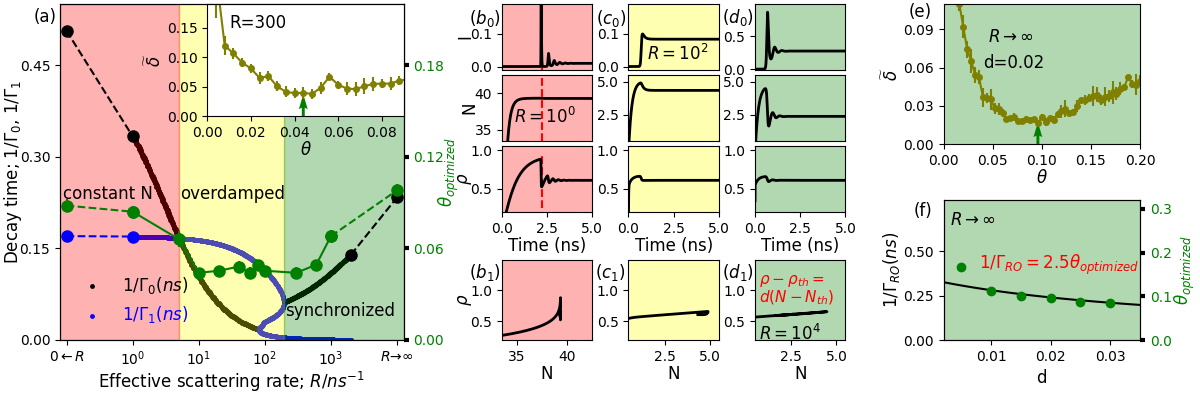}
\caption{(a) Decay times $1/\Gamma_{0}$ (black) and $1/\Gamma_{1}$ (blue) as a function of the effective scattering rate $R$ (determined from Eq.\eqref{eq:character}). Inset shows the RC error $\widetilde{\delta}$ for the QD laser with $R = 300$\,ns (Lorenz 63 task) as a function of the virtual node interval $\theta$ (green arrow points to $\theta_{opt}$). Green solid circles display $\theta_{opt}$ as a function of $R$. ($b_{0}$)-($d_{0}$) time-series of the intensity $I$, QW charge carrier density $N$, and QD charge carrier occupation probability $\rho$ for QD lasers with $R=10^{0}$ (orange shading), $R=10^{2}$ (yellow shading), and $R=10^{4}$ (green shading). ($b_{1}$)-($d_{1}$) $\rho-N$ phase diagrams. 
(e) $\theta$-scan of $\widetilde{\delta}$ for a QD laser with $ R \rightarrow \infty$ and $d=0.02$. (f) RO decay time $1/\Gamma_{RO}$ (black line) and $\theta_{opt}$ (green solid circles) for QD lasers with $R \rightarrow \infty$ as a function of the detailed balance coefficient $d$. Parameters as given in  Tab.\ref{Tab:QDparameters} and \ref{Tab: CW-NRMSE}. } \label{fig: Regiems_1} 
\end{figure*}

The most relevant parameters of the delay-based QD laser RC setup, excluding the optical injection which will be discussed later in Section \ref{sec:inj}, are the delay time $\tau$, the feedback phase $C$, and the feedback strength $K$. For a given QD laser (with a fixed value of $R$) with a constant $\tau$, we initially scan the emission dynamics within the $(C, K)$-plane, without any RC input, to understand the baseline dynamical behavior of the QD laser system. The emission dynamics for the case $R=1$ are explored in Fig.\ref{fig: figurenew}a-c using three different initial conditions$:$ up sweep in $C$ (a), down sweep in $C$ (b), and constant initial conditions (c). The intensity of the constant wave (cw) emission is color-coded using blue shading,  while oscillatory dynamics are colored according to the number of unique maxima, with yellow areas representing regions with more than 4 maxima. As known for semiconductor lasers with delayed feedback \cite{ROT07, LUE20} the 
$(C, K)$-plane is structured by saddle-node (SN) bifurcations (cyan lines in Fig.\ref{fig: figurenew}) which depend solely on the field equation, and Hopf bifurcations (dashed lines in Fig.\ref{fig: figurenew}), which are significantly influenced by the scattering lifetime \cite{PET95a, LEV95, LUE11, GLO12, OTT11}. 
The different sweeps in Fig.\ref{fig: figurenew} show multi-stability which arises due to new stable solutions being born at each SN bifurcation and, depending on the sweep direction, the system follows a different cw solution branch. The two SN lines meet at a cusp bifurcation \cite{DIE20}. For large values of the feedback strength $K$ the first stable solution loses its stability through a Hopf-bifurcation  (represented by black dashed lines) and complex emission dynamics are observed beyond that line (red/yellow area in Fig.\ref{fig: figurenew}). Similar bifurcation scenarios occur for other $R$ values. The main difference being the position of the Hopf-bifurcation, which is shifted along the $K$-axis depending on $R$, as will be discussed later.



The RC performances presented later on are calculated using constant initial conditions, for which the solutions settle to either the up-sweep or the down-sweep solution branches. Fig.\ref{fig: figurenew}c shows the dynamics for the initial conditions that will be used in the subsequent results, which are a combination of what is observed in Fig.\ref{fig: figurenew}a and b. 
An example of the Lorenz 63 prediction task performance in $(C-K)$- plane for $R=1$ is shown in Fig.\ref{fig: figurenew}d. In comparison with Fig.\ref{fig: figurenew}c, we see that the dynamics greatly influence the performance. For this task, good performance (low errors) is obtained in the cw regions far from the Hopf bifurcation. Beyond the Hopf bifurcation, the RC performance deteriorates significantly (white areas). 

\subsection{Impact of QD laser dynamics}\label{the performance compare}

The response timescales of the internal QD laser dynamics are crucial for understanding the system's memory and processing capabilities. To explore the effect of the internal timescales on the computing properties of the RC system, we analyze the decay times $1/\Gamma_{0}$ and $1/\Gamma_{1}$ of the solitary QD laser system, determined from Eq.\eqref{eq:character}, as a function of $R$ (black and blue lines Fig.\ref{fig: Regiems_1}a). Three distinct dynamical regimes can be identified: the constant QW charge carrier density regime (pink shading), the overdamped regime (yellow shading), and the synchronized regime (green shading) \cite{LIN14, LIN12,lin15a}. 

These regimes differ in their turn-on dynamics, which are visualized in Fig.\ref{fig: Regiems_1}b$_0$-d$_0$ showing the three dynamic variables: intensity $I$, QD-occupation $\rho$, and non-resonant carrier number $N$. Phase portraits are plotted in the Fig.\ref{fig: Regiems_1}b$_1$-d$_1$
The red dashed lines in Fig.\ref{fig: Regiems_1}b$_0$ highlight the laser's turn-on point, emphasizing that in the constant QW charge carrier density regime, the QW charge carrier density $N$ remains constant both before and after the laser turns on. In this regime, only the QD carriers react to the changes in laser intensity. 
In contrast, within the overdamped regime  in Fig.\ref{fig: Regiems_1}c$_0$, the optical intensity $I$ exhibits fast decay with strong damping and without any relaxation oscillations. In the synchronized regime, the phase diagram  in Fig.\ref{fig: Regiems_1}d$_1$ shows an approximately linear relationship between $\rho$ and $N$, indicating synchronization which is given by the static relation $\rho-\rho_{th}=d(N-N_{th})$.
Further details on the QD laser dynamics across these regimes can be found in the supplementary material.

The interplay between the internal timescales of the QD lasers and the time scales on which data is fed into the laser can crucially influence the performance. This is because 
$\theta$ determines whether next-neighbor interactions between virtual nodes take place and what the effective nonlinear transform of the input data is. If $\theta$ is much longer than the characteristic decay time of the system, there is no interaction between neighboring virtual nodes and the internal dynamics decay fully before the subsequent input arrives. In such cases, the system can be simplified to an iterative map \cite{HUE22, JAU24}, where the only coupling between nodes is introduced through the external delay line. However, if $\theta$ is too small, the laser does not have enough time to react to the input, which hinders a consistent and complete response. Since $\theta$ determines the data processing rate ($1/T=1/N_v\theta$), we are interested in how the optimal $\theta$ varies in dependence of the internal QD laser dynamics.
The optimal choice of $\theta$ is however also dependent on the requirements of the specific task. Fig.\ref{fig: Regiems_1}a shows how the optimal node distance for the Lorenz task prediction performance changes if the QD laser is changed via $R$ (green circles showing the values obtained via $\theta$-scans, as exemplarily displayed in the inset of Fig.\ref{fig: Regiems_1}a and in Fig.\ref{fig: Regiems_1}e).  The optimal $\theta$ closely follows the variations in the decay times $1/\Gamma_{0}$ and $1/\Gamma_{1}$, (black and blue lines) which nicely visualizes how the optimal input timescale is influenced by the system response time. Here, we have first restricted our study to a fixed feedback delay time and have optimized the feedback phase $C$, feedback strength $K$, data injection strength $\eta$ and mask length $\theta$. See the supplemental material for a description of the optimization procedure.
The optimized parameter values are summarized in column 2-5 of Tab.\ref{Tab: CW-NRMSE} for each QD laser (each $R$).


For the QD laser with instantaneous scattering ($R\rightarrow \infty$), the system behaves like a QW laser and the dynamics of the occupation probability of the quantum dots $\rho$ can be adiabatically eliminated (see the supplementary material for details). The reduced system is characterized by only one pair of complex eigenvalues that determines the decay time $1/\Gamma_{RO}=1/\Gamma_{0}$. Fig.\ref{fig: Regiems_1}f shows the optimal $\theta$ and the decay time, in this QW limit, as a function of the detailed balance coefficient $d$. A ratio of 2.5 between both is found which 
,however, strongly depends on other system parameters as e.g. delay times $\tau$, number of virtual nodes $N_{V}$, and task, as will be shown later.

\begin{table*}[hb]
\caption{Optimized parameters for QD reservoir for Lorenz task. $\eta_{opt}$ and $\theta_{opt}$ are determined for $\tau=1.692 \ ns$ via iterative optimization. $\eta_{2D}$ and $\theta_{2D}$ are chosen for 2D performance scans, $\theta^{\tau T}_{opt}$ and $\tau^{\tau T}_{opt}$ are parameters where best performance is found in $(\tau,T)$-plane.}
\begin{threeparttable}
\begin{tabular}{c c c c c| c c | c c } 
\headrow
$R$/ns$^{-1}$ & $K_{opt}$ & $C_{opt}/\pi$ & $\eta_{opt}$ & $\theta_{opt}/$ns & \hspace{2mm}$\eta_{2D}$ \hspace{2mm}& $\theta_{2D}/$ns & \hspace{1mm}  $\theta^{\tau T}_{opt}/$ns\hspace{1mm} &\hspace{1mm}$\tau^{\tau T}_{opt}/$ns\hspace{1mm} \vspace{1mm}\\ 
 0  &  0.018 & 0.72 & 0.5 & 0.088 & 0.5 & 0.082 & 0.073 & 1.36 \\
 1  &  0.024 & 0.36 &  0.4 & 0.084 & 0.6 & 0.084  &0.062 &  1.08\\
  2.7&        &      &     &       & 0.7 & 0.072  & & \\
 5  &  0.01  & 1.2 & 0.8 & 0.066 & 0.8 & 0.052 & & \\
 10 & 0.012 & 1.64 & 0.8 & 0.044  & 1.2 & 0.044  &0.059 & 2.32 \\
 20 & 0.008 & 1.5 & 1 & 0.045   & 1.4 & 0.045  &0.064 &2.64 \\
 40 & 0.011 & 1.4 & 1.1& 0.048    & 1.5& 0.048 & 0.064& 2.12\\
 60 & 0.008 & 1.64 &  1.1& 0.044   & 1.6& 0.044  & 0.052& 2.28 \\
 80 & 0.015 & 1.42 & 1.1& 0.049    & 1.2& 0.049 & 0.056& 2.19\\
 100 & 0.007 & 1.62 & 1.1& 0.045  & 1.8& 0.045     & 0.057& 2.28\\
 300 & 0.004 & 1.22 &  1.3& 0.044    & 1.6& 0.044  &0.06 & 2.44\\
 600 & 0.015 & 1.46 &  1.3 & 0.049    & 1.6 & 0.049     & 0.057& 2.12\\
 1000 & 0.01 & 1.22  & 1 & 0.068    &0.9 & 0.068 & 0.087& 1.84\\
 $\infty$ & 0.014 & 1.18 &  0.6 & 0.098   &0.6 & 0.08 & 0.1&   2\\
\hline 
\end{tabular}
\end{threeparttable}
\label{Tab: CW-NRMSE} 
\end{table*}

\begin{figure}[t]
\includegraphics[width=0.49\textwidth]{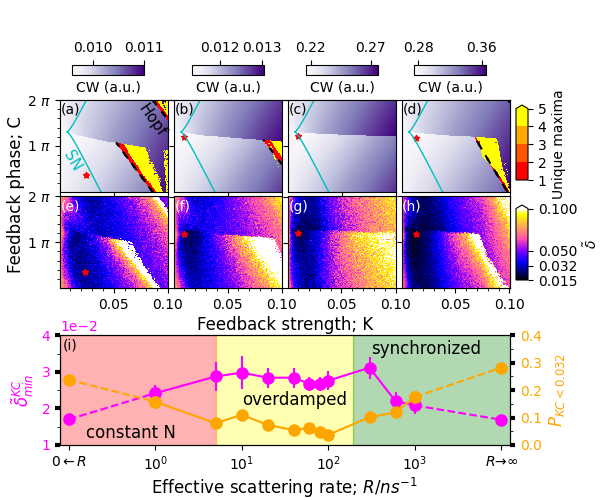}
\caption{\label{fig: CW_NRMSE} 
2D parameter scans in the $(K, C)$-plane for the Lorenz 63 one-step ahead prediction task for QD laser reservoirs with feedback: (a-d) dynamics without input and (e-h) prediction performance for  $R=1$, $R=2.7$, $R=10^{3}$, and $R \rightarrow \infty$. Stars mark the best performance $\tilde{\delta}_{min}^{KC}$ for each QD laser. (i) Best performance $\tilde{\delta}_{min}^{KC}$ (magenta symbols) and sensitivity $P_{KC<0.032}$ (orange symbols) as a function of $R$. Parameters:  $\tau=1.602 \ ns$ and as given in Tab.\ref{Tab: CW-NRMSE}.
} 
\end{figure}

\subsection{Performance for Lorentz 63 task}\label{Lorenz}
We now aim to investigate which of the different QD lasers exhibits the overall best performance. Based on the more complex charge carrier dynamics of QD lasers (higher dimensional phase space), it was initially expected that QD lasers would perform better than their QW counterparts that only have one charge carrier type. However, our findings indicate that pronounced ROs in the laser turn-on are beneficial for the performance even if the phase space is of higher dimension. Nevertheless, the QDs with the pronounced ROs require slower injection rates (larger $\theta$). 
 
The 2D bifurcation diagrams in the $(C, K)$-plane shown in Fig.\ref{fig: CW_NRMSE}a-d provide insight into how the dynamics of the QD laser system evolve as R increases. As discussed earlier in Fig.\ref{fig: figurenew}, complex dynamics emerge beyond the Hopf-bifurcation (dashed line). Due to the larger turn-on damping for the intermediate R-values, the Hopf bifurcation shifts to higher values of $K$\cite{KAN05, GLO12, OTT12} and thus outside of our scanning region. In contrast, as detailed in the supporting material, the SN line remains unaffected by variations in $R$. 
Comparing the scans of the dynamics with the corresponding RC performance (Fig.\ref{fig: CW_NRMSE} upper and lower row), we find that across all QD lasers (for all $R$ values), poor performance (large $\widetilde{\delta}$) occurs in the regions exhibiting complex dynamics. Instead, good performance (small $\widetilde{\delta}$) is observed within the cw regions at small $K$. The best values $\widetilde{\delta}_{min}^{KC}$ are marked by red stars in Fig.\ref{fig: CW_NRMSE}e-h with its value plotted with magenta symbols in Fig.\ref{fig: CW_NRMSE}i. Interestingly, optimal performance is observed for both very large and very small $R$ values, while the intermediate range of $R$, corresponding to the overdamped regime of QD lasers, performs less favorably. 
This trend is further supported by the sensitivity $P_{KC<0.032}$, depicted by the orange symbols in Fig.\ref{fig: CW_NRMSE}i.

\begin{figure}[t]
\includegraphics[width=0.48\textwidth]{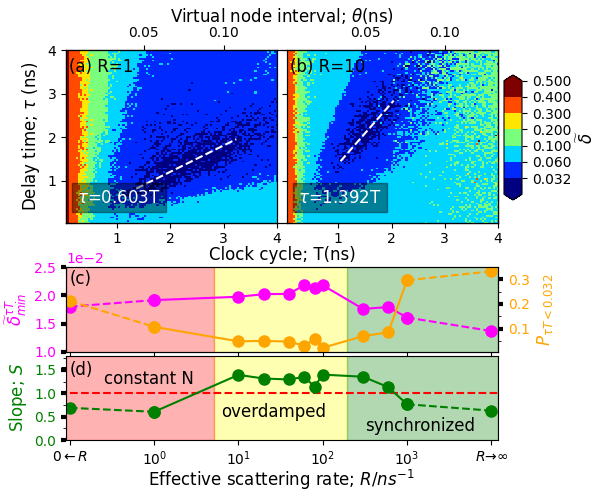}
\caption{\label{fig: T_tau} Lorenz task performance in $(\tau,T)$- plane: (a) and (b) show 2D performance scans  for two QD laser reservoirs ($R=1$; $R=10$) under optical feedback, top x-axis gives $\theta$-values via $T$=$N_{v} \theta$, color encodes the prediction error $\widetilde{\delta}$, white dashed lines show linear fits of the best-performing points ($\widetilde{\delta}<1.2 \widetilde{\delta}_{min}^{\tau T}$). (c) Best performance $\widetilde{\delta}_{min}^{\tau T}$ (magenta) and sensitivity $P_{\tau T<0.02}$ (orange) as a function of the scattering rate $R$, (d) Green symbols show the fitted slopes $S=\tau/T$, dashed red line indicates $\tau/T$=1. $N_{v}=30$, other parameters and optimized values are given in Tab.\ref{Tab: CW-NRMSE}.
} 
\end{figure}

Two things are interesting to note here: First, the QD lasers with overdamped dynamics are less favorable for a good performance and second, the best 
 performance is not found close to the Hopf-bifurcation as is often stated \cite{YAN22, FOL19, HOU19a, DAN23, LI23b}.  
 One reason is that the Lorenz task has a  lower memory requirement and does not need the strong undamping of the ROs close to the Hopf-bifurcation.
The memory of the RC setup is influenced by two key factors: the virtual node interval $\theta$, which governs the interaction between neighboring virtual nodes, and the delay time $\tau$, which determines the coupling across different clock cycles. Conducting a performance scan in the $(\tau, T)$-plane shows how the optimal $T$ shifts with the delay to best utilize both contributions to the memory.
For two selected QD lasers we show $(\tau,T)$ scans in Fig.\ref{fig: T_tau}a,b ($\theta$ is given by the top x-axis via $T=N_{v} \theta$). 
The points with the best performance in the $(\tau, T)$-plane are scattered around  straight lines (white dashed lines). We determine the white lines via fitting the points with prediction errors $\widetilde{\delta}<1.2\widetilde{\delta}_{min}^{\tau T}$ ($\widetilde{\delta}_{min}^{\tau T}$ is the best performance observed in the plane). This highlights that for each given delay time $\tau$, there is a different optimal virtual node distance $\theta$ (optimal clock cycle $T$). From the last section we know that $\theta$ is strongly influenced by the internal decay times. However, now we also find that the optimal ratio between both is not fixed but dependent on the delay time $\tau$.  The slopes $S=\tau/T$, are plotted as a function of $R$ in green in Fig.\ref{fig: T_tau}d (see supplementary material for all $\tau- T$  scans).

Surprisingly the slope $S$ is largest in the overdamped region with a value around 1.4. As can be seen in Fig.\ref{fig: T_tau}d, $S>1$ in the overdamped regime, while $S<1$ in the other two regimes. This indicates that the slope $S$ is determined by the dynamics of the reservoir. QD lasers with pronounced turn-on oscillations $\Gamma_{RO}<\omega_{RO}$ favor that one data point is injected longer than one delay time. Instead, strongly damped responses perform better when more than one data point is injected during one delay. It is noted that changing the number of virtual nodes $N_{v}$ will also change $S$ (see supplementary material). The most drastic impact on the slope is given by the task.  As discussed next, the optimal slope is around 2 for the 2-step ahead Mackey-Glass task. 

The best performance $\widetilde{\delta}_{min}^{\tau T}$ (magenta) and sensitivity $P_{\tau T <0.032}$ (orange) are plotted as a function of $R$ in Fig.\ref{fig: T_tau}c. Similar to the dependence of the best performance on $R$ when scanning $C$ and $K$ (Fig.\ref{fig: CW_NRMSE}i), we find that QD lasers with very large $R$ perform better. We also find that optimized values for $\tau$ increase the performance with the best value $\widetilde{\delta}_{min}^{\tau T}=0.0137$ found for QD lasers that resemble quantum well lasers ($R \rightarrow \infty$) for $\tau$=2\,ns and $\theta$=100\,ps.

\subsection{Performance  Mackey-Glass task}

\begin{figure}[t]
\includegraphics[width=0.48\textwidth]{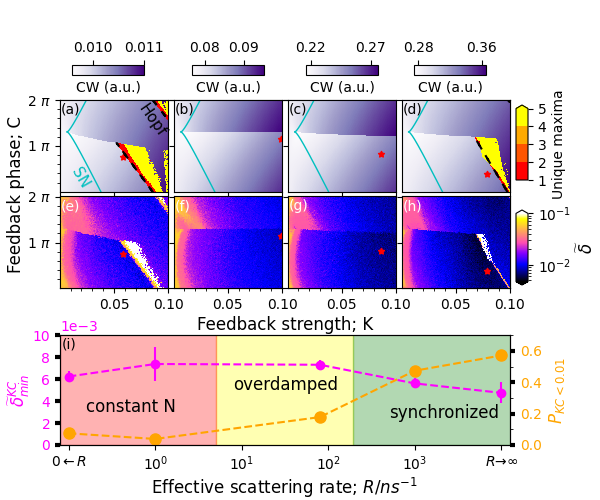}
\caption{\label{fig: CW_NRMSE_Mackey_glass} 
2D parameter scans in the $(C, K)$-plane for the Mackey-Glass two-step ahead prediction task plotted for four QD laser reservoirs  with optical feedback (compare  Fig.\ref{fig: CW_NRMSE}). (a-d) dynamics without input, and (e-h) performance for ($R$=1; $R$=80; $R$=$10^{3}$; $R\rightarrow\infty$). Stars mark best performance $\tilde{\delta}_{min}^{KC}$ for each QD laser RC. (i)Best performance $\tilde{\delta}_{min}^{KC}$ (magenta symbols) and sensitivity $P_{KC<0.01}$ (orange symbols) as a function of $R$. Parameters are given in Tab.\ref{Tab:Mackey-Glass}, $\tau=1.692 \ ns$.}
\end{figure}

\begin{figure}[t]
\includegraphics[width=0.49\textwidth]{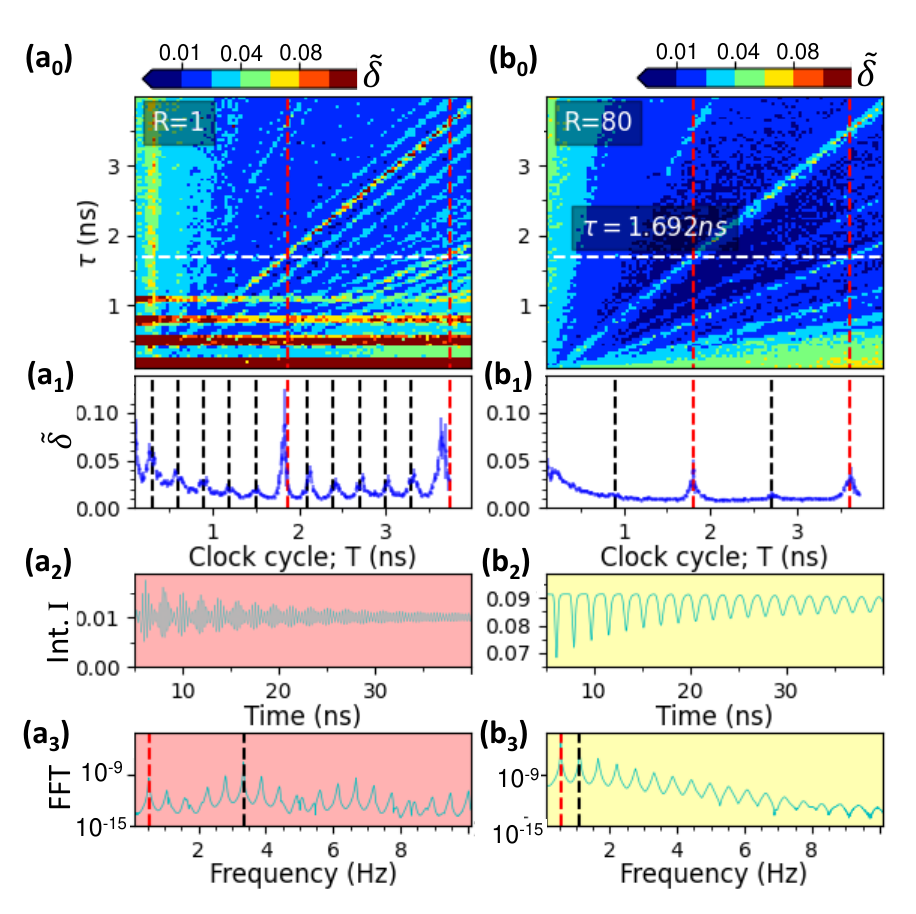}
\caption{\label{fig: T_tau_resonance_mackey_Glass} (a$_0$,b$_0$)$:$ 2D performance scans in the $(\tau, T)$-plane for the Mackey-Glass two-step ahead prediction task for two QD laser reservoirs ($R=1$; $R=80$) under optical feedback. White dashed lines indicate line scans in (a$_1$,b$_1$) at $\tau=1.692 \ ns$. (a$_2$,b$_2$) show the time series of the QD laser intensities $I$ and (a$_3$,b$_3$) show their Fast Fourier transforms. Red and black dashed lines in (a$_3$,b$_3$) indicate $f_0 \approx 1/\tau$ and the main frequencies $f_1$ (highest peak). In (a$_0$,a$_1$) and (b$_0$,b$_1$), dashed lines mark the multiples of $1/f_0$ and $1/f_1$. Parameters are given in Tab.\ref{Tab:Mackey-Glass}. }
\end{figure}


We will now conduct our analysis using a 2-step ahead Mackey-Glass task (see Sec.\ref{Tasks}). This task is expected to require more memory compared to the Lorenz 63 task, allowing us to assess the extent to which our previous conclusions are task-dependent (we will use the same delay-based QD laser reservoirs).
In Fig.\ref{fig: CW_NRMSE_Mackey_glass}, we present the 2D performance scans in $(C,K)$-parameter plane, similar to the approach taken in Fig.\ref{fig: CW_NRMSE}. Fig.\ref{fig: CW_NRMSE_Mackey_glass}a-d illustrate again the dynamics of each QD laser without input, and Fig.\ref{fig: CW_NRMSE_Mackey_glass}e-h depict the performances of the corresponding RC setup. Stars mark the positions where the best performance $\tilde{\delta}_{min}^{KC}$ is found for each QD laser reservoir. Unlike the performance diagrams observed for the Lorenz task (Fig.\ref{fig: CW_NRMSE}e-h), the optimal performance zone for the Mackey-Glass task is located closer to the Hopf-bifurcation line at a position where the feedback already leads to a strong undamping of the turn-on transients\cite{ERN10b, LUE20}. The optimal $K$ values (the positions of the red stars) are approximately a factor of 8 larger than the optimal $K$ values for the Lorenz task. This can be attributed to the increased memory requirement of the Mackey Glass task\cite{JAU24}. In contrast to Lorenz, the Mackey-Glass time series exhibits long-term correlations, necessitating the reservoir to retain past inputs for a significantly longer duration.  The slow damping induced by the high feedback strength allows the reservoir to maintain longer memory of past states, which is crucial for accurately predicting the Mackey-Glass time series.


 

In Fig.\ref{fig: CW_NRMSE_Mackey_glass}i we depict the optimal Mackey-Glass prediction performance found in the 2D scans as a function of the effective scattering rate $R$ (magenta symbols).  The best performance $\widetilde{\delta}_{min}^{KC}$ and sensitivity $P_{KC<0.01}$ (orange symbols) are found for very small and very large values of $R$ while the typical QD laser with $R=100$ is slightly worse by a factor of $1.4$. This trend is similar to what was observed for the Lorenz task but  not as pronounced. This is expected, as the next neighbor coupling, which is mostly impacted by the local dynamics and thus by $R$, is not the most important contribution here, rather, the delay induced contribution dominates.    

\begin{figure}[t]
\includegraphics[width=0.48\textwidth]{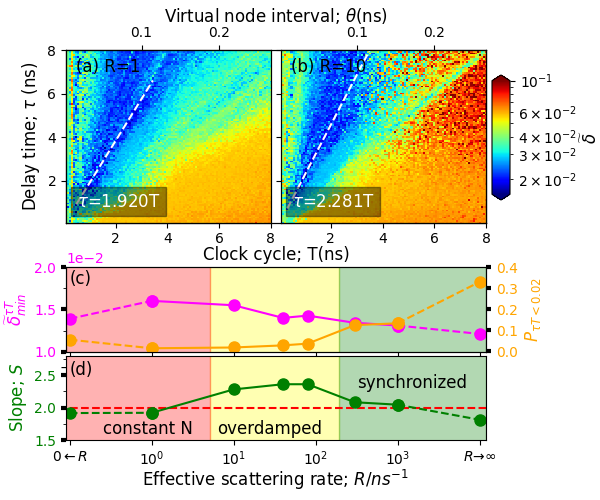}
\caption{\label{fig: T_Tau_Mackey_Glass_fixedK0_01_C1.5} 
Mackey-Glass two-step ahead prediction performance in $(\tau, T)$-plane$:$ (a) and (b) show 2D performance scans for QD lasers with $R=1$ and $R=10$, respectively, under optical feedback with $(K, C)=(0.01,1.5 \pi)$, color encodes the prediction error $\widetilde{\delta}$,  white dashed lines show linear fits of the best-performing points ($\widetilde{\delta}<1.2 \widetilde{\delta}_{min}^{\tau T}$)(compare Fig.\ref{fig: T_tau}). (c) Best performance $\widetilde{\delta}_{min}^{\tau T}$ (magenta) and sensitivity $P_{\tau T<0.02}$ with $\widetilde{\delta}_{threshold}=0.02$ (orange) as a function of the scattering rate $R$, (d) Green symbols show the fitted slopes $S=\tau/T$, dashed red line indicates $\tau/T$=2. $N_{v}=30$, other parameters are given in Tab.\ref{Tab:Mackey-Glass}.
} 
\end{figure}

To clarify how the performance depends on $\tau$ and $T$ for the Mackey-Glass task, we performed performance scans in the $(\tau, T)$-plane, which are displayed in Fig.\ref{fig: T_tau_resonance_mackey_Glass}a$_0$,b$_0$.  Unlike the results for the Lorenz task (Fig.\ref{fig: T_tau}), we find very pronounced resonances with deteriorated performance at $m \tau=n T$ ($m,n \in \mathbb{N}$) and regions where the dynamics are complex and RC becomes unfeasible (red colors).
The line scans along $T$ in Fig.\ref{fig: T_tau_resonance_mackey_Glass}a$_1$,b$_1$ are taken for a fixed $\tau$ (horizontal white line in Fig.\ref{fig: T_tau_resonance_mackey_Glass}a$_0$-b$_0$)  and clearly show the pronounced peaks in the computing error. These peaks emerge because the chosen optimal $(K, C)$ values for the Mackey-Glass task (marked by red stars in Fig.\ref{fig: CW_NRMSE_Mackey_glass}) are very close to the Hopf-bifurcation line. Operating near this bifurcation means that the delay-based QD laser reservoir is close to its instability threshold and therefore susceptible to input perturbations. In the case of $R=1$ (shown in Fig.\ref{fig: T_tau_resonance_mackey_Glass}a$_0$), we also observe that even slight changes in $\tau$ can  push the reservoir in and out of its stable operating region (horizontal red regions).

\begin{table*}[hb]
\caption{Optimized parameters for the QD reservoir with Mackey-Glass two-step prediction. $\eta_{opt}$ and $\theta_{opt}$ are determined for $\tau=1.692 \ ns$ via iterative optimization. $\eta_{2D}$ and $\theta_{2D}$ are used for $(K,C)$ performance scans. Last three columns with $\eta_{fixed}$, $\theta_{fixed}$, $\tau_{fixed}$ are optimal values after $(\tau,T)$-scans with $K=0.01$,$C=1.5 \pi$.} 
\begin{threeparttable}
\begin{tabular}{c c c c c |c c|c c c} 
\headrow
$R$/ns$^{-1}$ & $K_{opt}$ & $C_{opt}/\pi$ & $\eta_{opt}$ & $\theta_{opt}/$ns & $\eta_{2D}$ & $\theta_{2D}/$ns & $\eta_{fixed}$ & $\theta_{fixed}$ & $\tau_{fixed}$\\
 0        &   0.077   &  0.70  & 0.18    &  0.068  &  0.23 & 0.047  & 0.5 &0.061&2.96 \\
 1         & 0.058  &  0.76  & 0.39  & 0.066 &0.40  & 0.066   &0.5 &0.053&3.52\\
 10     &      &         &      &       &     &      & 0.9 &0.04 &2.8\\
 40      &      &         &      &       &     &      & 1.1 &0.032 & 2.15\\
 80     & 0.099   &  1.14  & 0.70   & 0.049& 0.50  & 0.070 & 1.2  &0.027&1.84\\
 300    &      &         &      &       &     &      & 1.0 & 0.053 & 3.36\\
 1000    & 0.086  &  0.82     &  0.50 & 0.069&  0.50  &0.069 &1.0 & 0.04&2.8\\
 $\infty$  & 0.079   &  0.38  & 0.45 & 0.067& 0.30  & 0.068 & 0.60 & 0.067 & 4.56\\
\hline 
\end{tabular}
\end{threeparttable}
\label{Tab:Mackey-Glass} 
\end{table*}

The peaks in the error occur for data injection rates $1/T$ that are multiples of the main resonance frequency of the reservoir. To illustrate that we plot the transient response  and its Fourier transform (FFT) of the QD laser with feedback but without input in the last two rows of Fig.\ref{fig: T_tau_resonance_mackey_Glass}. Panel (a3,b3) show the FFT with the red dashed lines indicating $f_0 \approx 1/\tau$, and the black dashed lines indicating the main frequency $f_1$ (highest intensity excluding $f_0$). In Fig.\ref{fig: T_tau_resonance_mackey_Glass}a$_0$,b$_0$ and a$_1$,b$_1$ the red and black dashed lines mark  $n/f_0$, and $n/f_1$ ($n \in \mathbb{N}$). The resonance peaks in the error align with these frequencies indicating that they are directly tied to the intrinsic oscillatory dynamics of the system. 

Comparing Fig.\ref{fig: T_tau_resonance_mackey_Glass}a and (b), we observe that the QD laser with strong damping ($R=80$) exhibits more consistent performance in the $(\tau,T)$-parameter plane. This difference arises mainly from the proximity of the optimal feedback strength $K$ to the Hopf-bifurcation. For the QD laser with pronounced ROs ($R=1$), the optimal $K$ lies closer to the Hopf-bifurcation, while for the QD laser with strong damping ($R=80$), the Hopf-bifurcation occurs only at much higher $K$ values, well beyond the optimal $K$. Therefore, for practical applications, a typical QD laser with strong turn-on damping is slightly worse in performance but more resilient to parameter variations.

As we have observed so far, the optimal feedback parameters $(K,C)$ for the Mackey-Glass task are not well suited for scans in the delay time due to the proximity of the Hopf-bifurcation. We therefore perform this analysis with significantly smaller $K$ values, i.e. with much larger distance to the Hopf bifurcation at $(K,C)=(0.01,1.5 \pi)$. Fig.\ref{fig: T_Tau_Mackey_Glass_fixedK0_01_C1.5}a,b, show $(\tau,T)$-scans for two different QD lasers. The resonance regions with very bad performance disappeared and instead a cone of good performance regions are found (blue points) similar to what we have seen for the Lorenz task. 
As done before we determine the slopes $S=\tau/T$ via fitting the best-performing region with $\widetilde{\delta}<1.2\widetilde{\delta}_{min}^{\tau T}$ and plot it as a function of $R$ in Fig.\ref{fig: T_Tau_Mackey_Glass_fixedK0_01_C1.5}d. 
The best performance $\widetilde{\delta}_{min}^{\tau T}$ (magenta) and the sensitivity $P_{\tau T <0.02}$ (orange) are shown in Fig.\ref{fig: T_Tau_Mackey_Glass_fixedK0_01_C1.5}c. We observe that QD lasers with large $R$ values perform best, similar to what we learned for the Lorenz task. Nevertheless the maximum error (worse performance) shifted to smaller $R$. Note that due to the non-optimal feedback strength, the best performance decreased by a factor of 2 compared with Fig.\ref{fig: CW_NRMSE_Mackey_glass}. 

The ratio between the delay and the clockcycle, the slope $S$, now varies around 2, again with the highest value reached for the strongest damping (Fig.\ref{fig: T_Tau_Mackey_Glass_fixedK0_01_C1.5}d).  This indicates that for this Mackey-Glass task it is better to feed two inputs into one delay loop, which is related to the higher memory demand of the task. 
It appears that the analysis of the $\tau/T$ slope can serve as an effective method for evaluating the memory requirements of different tasks. 
It further provides valuable insight into the optimal configuration of reservoir computing systems, guiding the selection of parameters to achieve the best performance for various prediction tasks with differing memory needs.

\subsection{RC with optical injection}\label{sec:inj}

External optical injection has been employed to enhance the performance and processing speed of RC systems based on semiconductor lasers \cite{YAN22, EST23, EST20}. To complete our analysis and explore further ways to improve performance, we now apply additional external optical injection to the RC setup we have discussed so far. Although data are still injected via the electric pump current, the addition of optical injection allows for the fine-tuning of the system's response time and underlying dynamics.
The complete electric field equation for the QD laser with optical injection is described by Eq.\eqref{eq:QDL_injected_feedback}, with the key difference that the $K_{inj}$ is now non-zero.

\begin{figure}[t]
\includegraphics[width=0.48\textwidth]{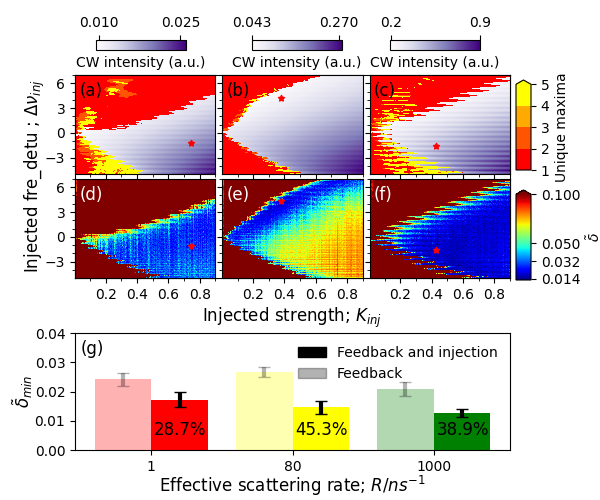}
\caption{\label{fig: Kinj_deta_nu} 
2D parameter scans in the $(K_{inj}$, $\Delta \nu_{inj})$-plane for the Lorenz 63 one-step ahead prediction task are presented for three QD laser reservoirs ($R$=1; $R$=80; $R$=$10^{3}$) under optical injection and feedback. (a-c) illustrate the dynamics of each QD laser without input, while (d-f) depict the performances of the corresponding RCs.  (g) Best computing errors achieved under optical injection, $\tilde{\delta}_{min}^{K_{inj} \Delta \nu_{inj}}$, marked by stars in the $(K_{inj}$, $\Delta \nu_{inj})$-plane (dark bars), while the reference performances without optical injection (from Fig.\ref{fig: CW_NRMSE}i), are shown using light bars. Numbers in dark bars give improvement values. Parameters are given in Tab.\ref{Tab: injection}. 
} 
\end{figure}

As expected for a laser with optical injection \cite{WIE05, PAU12, LIN14}, we observe a locking cone in the parameter space of the optical injection, i.e. the ($K_{inj}$, $\Delta \nu_{inj}$)-plane. In Fig.\ref{fig: Kinj_deta_nu}(a-c), we show 2D bifurcation diagrams for three QD lasers with different scattering rates$:$ $R$ = 1, $R$ = 80, and $R$ = 1000. The steady-state intensities are depicted with purple shading, while the number of unique maxima is represented as before by red to yellow colors as indicated in the legend. Given the presence of both optical injection and optical self-feedback, the locking boundary is modulated by small resonances caused by the external cavity modes induced by the delay \cite{MUE24, KOE21b}. In Fig.\ref{fig: Kinj_deta_nu}d-f, we present the 2D performance scans in the same parameter plane. One notable observation is that acceptable performance is only achieved within the locking region; outside of this region, the laser enters a regime of complex dynamics that is not suitable for reservoir computing. The upper boundary of the locking region at positive detunings is highly dependent on the chosen $R$, as it is given by a Hopf-bifurcation that shifts with the relaxation RO damping \cite{PAU12}. Furthermore, the optimal operational conditions in the ($K_{inj}$-$\Delta \nu_{inj}$) plane are influenced by the turn-on damping characteristics of the solitary QD laser. Specifically, the QD laser with overdamped dynamics (Fig.\ref{fig: Kinj_deta_nu}e) operates best near the locking boundary at positive detunings. In contrast, QD lasers exhibiting pronounced ROs (Fig.\ref{fig: Kinj_deta_nu}f) perform well over a broader range of parameters.

\begin{table}[b]
\caption{Parameter for $(K_{inj}$-$\Delta \nu_{inj})$ scans of Lorenz prediction performance for QD laser with feedback and injection.} 
\begin{threeparttable}
\begin{tabular}{c c c c c c} 
\headrow
\hspace{-2mm}$R/ns^{-1}$ &\hspace{-2mm} $K_{opt}$ &\hspace{-2mm} $C_{opt}/\pi$ & \hspace{-2mm}$\tau/$ns & $\theta_{opt}/$ns &\hspace{-2mm} $\eta_{opt}$  \\ [0.5ex]
\hline %
 1  & 0.031 & 1.28 & 1.692 & 0.048  &  2.0  \\
 80 & 0.020 & 0.72  & 1.692 &  0.064  &  1.3   \\
 1000 & 0.071 & 0.70 & 1.692  & 0.088   & 2.3  \\
[0.5ex]
\hline 
\end{tabular}
\end{threeparttable}
\label{Tab: injection} 
\end{table}

To quantify the improvement, we plot the best performance values $\tilde{\delta}_{min}^{K{inj} \Delta \nu_{inj}}$ in a bar diagram, as shown in Fig.\ref{fig: Kinj_deta_nu}g, alongside the reference performance $\tilde{\delta}_{min}^{KC}$ (from Fig.\ref{fig: CW_NRMSE}i), obtained without optical injection. For all QD lasers considered,  we observe a significant performance enhancement of more than 28\%. Notably, for the QD laser with overdamped dynamics, the improvement is particularly pronounced, reaching up to 45\%. This demonstrates that external optical injection is a highly effective method for enhancing reservoir computing performance, especially in laser systems with overdamped turn-on dynamics.

\section{Conclusion}\label{sec:conclusion}
We investigated the impact of internal scattering timescales within QD lasers on the computing performance if used as a delay-based time-multiplexed reservoir computer. We systematically studied the time series prediction performance when the scattering lifetimes are continously changed. Two different benchmark tasks, Lorenz 63 and Mackey-Glass, are used. The results reveal that better computing performance over a wider parameter range occurs for QD lasers that show pronounced relaxation oscillations during turn-on, e.g QD lasers with very fast scattering processes where they behave like conventional quantum well lasers. The typical QD laser, with its strongly damped response to perturbations, performs worse despite the fact that in this case the internal dynamics are of higher dimension. 

We find that the choice of an optimal input clock cycle, which determines the virtual node distance, impacts the performance. Further, we numerically show that the optimal virtual node distance is not only directly affected by the intrinsic decay time of the turn-on dynamics but also by the number of virtual nodes and the feedback delay time. We relate this to the effective coupling topology within the time-multiplexed reservoir setup. Both, the node distance and the delay change the internal coupling topology and it will change with different system dynamics. In our setup we observed this by monitoring the optimal ratio between delay and clock cycle which vary with the system response time. 

This being said, every time-series prediction task has different memory requirements, which is why the optimal setting for clock cycle and delay time need to be adjusted before operating a reservoir on a specific task. For tasks that need long memory to be accurately predicted, like the Mackey-Glass task, the effect of the delay to un-damped the relaxation oscillations can be utilized to increase the internal short term memory. However, in this case the close vicinity of the instability threshold given by the Hopf bifurcation leads to higher sensitivity to the system parameters.

Finally, we also showed that for all the QD lasers investigated here an additional external optical injection can strongly improve the reservoir computing performance by about 30\%.

In summary, we systematically studied the complex interaction between reservoir response time, input timescales and internal memory. We found that the underlying dynamics can be exploited for performance improvement, however, additional internal degrees of freedom of the reservoir dynamics is only useful if it occurs in the variable used for sampling the reservoir output. 

\section*{Acknowledgements}
KL acknowledges financial support by the European Union’s
Horizon 2020 programme under grant agreement number 101129904. LJ acknowledges funding from the Carl-Zeiss-Stiftung within the NEXUS program.

\section*{Conflict of interest}
the authors declare no conflict of interest

\section*{Supporting Information}

Supplementary material is provided via a pdf file that can be found via...

\printendnotes




\end{document}